\documentclass[aip, jcp, reprint,amsmath,amssymb]{revtex4-1}

\usepackage[pdftex]{graphicx}
\usepackage{dcolumn}
\usepackage{bm}
\usepackage{booktabs}
\usepackage{color}
\usepackage{url}
\usepackage{hyperref}
\makeatletter
\renewcommand{\@biblabel}[1]{\quad#1.}
\makeatother

\begin{document}
\title{Topological interactions between ring polymers: Implications for chromatin loops}
\author{Manfred Bohn}
\altaffiliation{The following article has been accepted by \textit{The Journal of Chemical Physics}. After it is published, it will be found at \url{http://jcp.aip.org}.}
\email{bohn@tphys.uni-heidelberg.de}
\author{Dieter W. Heermann}
\altaffiliation{Institute for Molecular Biophysics, The Jackson Laboratory, Bar Harbor, ME}
\altaffiliation{Interdisciplinary Center for Scientiﬁc Computing (IWR), University of Heidelberg, Im Neuenheimer Feld 368, D-69120 Heidelberg, Germany}
\affiliation{Institute for Theoretical Physics, University of Heidelberg, Philosophenweg 19, D-69120 Heidelberg, Germany}

\date{\today}
\begin{abstract}
Chromatin looping is a major epigenetic regulatory mechanism in higher eukaryotes. Besides its role in transcriptional regulation, chromatin loops have been proposed to play a pivotal role in the segregation of entire chromosomes. The detailed topological and entropic forces between loops still remain elusive. Here, we quantitatively determine the potential of mean force between the centers of mass of two ring polymers, i.e. loops.  We find that the transition from a linear to a ring polymer induces a strong increase in the entropic repulsion between these two polymers. On top, topological interactions such as the non-catenation constraint further reduce the number of accessible conformations of close-by ring polymers by about 50\%, resulting in an additional effective repulsion. Furthermore, the transition from linear to ring polymers displays changes in the conformational and structural properties of the system. In fact, ring polymers adopt a markedly more ordered and aligned state than linear ones. The forces and accompanying changes in shape and alignment between ring polymers suggest an important regulatory function of such a topology in biopolymers. We conjecture that dynamic loop formation in chromatin might act as a versatile control mechanism regulating and maintaining different local states of compaction and order.
\end{abstract}
\maketitle

\section{Introduction}
Ring polymers are abundantly found in nature, common examples range from the DNA of bacteria to certain virus types~\cite{Fiers1962}, mitochondrial DNA in higher eukaryotes and proteins~\cite{Zhou2003}. These findings suggest a structural advantage of ring polymers over linear ones.  The genomic content of higher eukaryotes inside the interphase nucleus, however, has a much more complex organization. Compaction is accomplished on several scales, the detailed folding pathways being closely related to genome function~\cite{Goetze2007, Luo2009}.  Nowadays, abundant experimental and theoretical evidence suggests that chromatin looping acts as an important link between chromatin organization and function. The existence of chromatin loops on scales from a few kilo basepairs (kb) up to several mega basepairs (Mb) has been evaluated by 3C/4C/5C experiments~\cite{Simonis2006, Lieberman-Aiden2009}. Several mechanisms for looping have been proposed: Amongst others, it was suggested that these loops arise when different genes and regulatory elements co-localize in transcription hubs or factories~\cite{Fraser2006}, where a high concentration of polymerases engages transcription.  

Inconsistent with theoretical predictions for solutions of linear polymers, chromosomes arrange such that they occupy distinct territories~\cite{Cremer2001}, the overlap volume being quite small. A similar effect has been found even for small parts of single chromosomes spanning  several mega basepairs (Mb)~\cite{Goetze2007}. In principle, active ATP-consuming mechanisms might be responsible for maintaining segregation. Rosa and Everaers~\cite{Rosa2008} proposed that chromosomes are aggregating into distinct compartments because of their large relaxation time, making it impossible for them to intermingle during the time of one cell cycle. However, this study does not explain segregation of smaller regions within chromosomes. Vettorel and co-workers~\cite{Vettorel2009} suggested that the collapsed state might be responsible for segregation:  Polymer gels, where single polymers are in a collapsed state, show no reptation, therefore intermingling between different polymers becomes practically impossible. Recently, evidence from polymer models has indicated that chromatin loops force such a compartmentalization already by virtue of purely entropic forces~\cite{Cook2009, Bohn2009b}.

While the mean squared radius of gyration $\left<R_g^2\right>$ of isolated ring polymers and a system of two catenated rings~\cite{Bohn2009c} displays a scaling similar to that of a self-avoiding walk, where $\left<R_g^2\right>\sim N^{2\nu}$ with $\nu\approx0.588$~\cite{Le1980}, non-catenated rings in a melt become compact adopting a scaling exponent of $\nu=1/3$~\cite{Mueller2000, Vettorel2009}. In fact, chromatin organization is not consistent with a simple melt of ring polymers. Scaling exponents found in experiments differ a lot, ranging from $\nu\approx 0.1-0.2$~\cite{Jhunjhunwala2008} to $\nu=0.5$~\cite{Engh1992}, the level of compaction depending on gene activity~\cite{Mateos2009}. Dynamic chromatin loops, however, have been shown recently to provide a consistent framework of chromatin organization~\cite{Bohn2009b}. 

The purpose of this study is to deepen the understanding of the topological interactions ring polymers exert on each other as well as the changes in their conformational properties induced by such forces. While topological effects of rings have been considered in several studies~\cite{Frank-Kamenetskii1975, Sikorski1994, Mueller1996}, the induced forces have not been quantitatively assessed. A system of two ring polymers can be viewed as a toy system for the influence of loop formation on chromatin folding. While the biological system is beyond doubt much more complicated -- multiple loops and the dense system of polymers playing a dominant role -- the effect of topological forces and the advantage of the ring shape for biopolymers can be highlighted best by studying it apart from other influences. 
We evaluate the potential of mean force between two ring polymers dependent on their mutual topological state to derive a quantitative measure for the entropic repulsion two rings or loops exert on each other. Therefore, Monte Carlo simulations of ring polymers with lengths of up to $N=2048$ monomers are conducted.  We separate the entropic effects due to excluded volume interactions from the topological interactions arising from the non-catenation constraint. The ring polymers' conformational properties and their mutual alignment subjected to the topological and entropic forces are investigated.  In fact, we find that ring polymers display a markedly stronger repulsion and adopt a more ordered state compared to linear polymers. The findings indicate a natural benefit of rings over linear ones, leading to the conjecture that chromatin loops not only facilitate entropy-driven segregation of chromosomes or intra-chromosomal regions, but act as a versatile control mechanism regulating and maintaining different states of compaction and order.

The paper is organized as follows: In section~\ref{sec:methods} we revisit the theory behind the effective potential, present details of the simulational method and the numerical calculations of the effective potential. In section~\ref{sec:results:saw} we calculate the potential of mean force for linear polymers and review scaling theories. We turn to ring polymers in section~\ref{sec:results:rings} and study the effect of topological interactions in section~\ref{sec:results:top}. The following two sections are dedicated to the investigation of how the dimensions and shape of a ring changes when bringing two rings into contact.  Finally, in section~\ref{sec:results:align}, we study the mutual alignment of rings considering the effect of topological constraints.

\section{Simulations and methods}\label{sec:methods}
\subsection{Effective potentials}
To investigate the  potential of mean force exerted between the centers of mass of ring polymers, we use the concept of the effective potential. The method has been used for a variety of polymer systems, and for some even analytical approximations have been found ~\cite{Dautenhahn1994, Hsu2004}. A recommendable review is given in Ref.~\cite{Likos2001}. 

Let the monomer positions and momenta of the two polymers be denoted by $\mathbf{r}_i^\alpha$ and $\mathbf{p}_i^\alpha \; (i=1, \ldots, N, \alpha=1, 2)$, respectively. $\alpha$ denotes the polymer chain index and $i$ indexes subsequently the monomers of the polymer. The properties of the system are given by the complete Hamiltonian 
\[ \mathcal{H}(\lbrace\mathbf{p}_i^\alpha\rbrace, \lbrace \mathbf{r}_i^\alpha \rbrace) = \sum_{\alpha=1,2}\sum_{i=1}^N\frac{{\mathbf{p}_i^\alpha}^2}{2m} + V(\left\lbrace \mathbf{r}_i^\alpha \right\rbrace) \]
For the problem of ring polymers investigated here, the interaction potential $V(\left\lbrace \mathbf{r}_i^\alpha \right\rbrace)$ is given by an excluded volume term and eventually by a term describing the topological constraints.

The potential of mean force can be gained via the partition sum of the system constrained to a fixed center-of-mass, 
\[ Z_c(\mathbf{R}) = \mathrm{Tr} \left[ \exp(-\beta \mathcal{H})\; \delta\left(\frac{1}{N}\sum_{i=1}^N \mathbf{r}_i^{(1)} - \frac{1}{N} \sum_{i=1}^N \mathbf{r}_i^{(2)} - \mathbf{R}\right)\right] \]
The trace denotes the integral over all momenta and monomer coordinates. 

The effective potential is now defined by the relation
\[ \exp\left(-\beta U_{\mathrm{eff}}(\mathbf{R})\right) = Z_c(\mathbf{R}) \]
For purposes of calculation, however, the following representation~\cite{Likos2001, Hsu2004} of the effective potential is more useful
\begin{equation}
\label{eq:effpot:simple}%
 U_{\mathrm{eff}}(R) = -k_B T \ln \frac{Z_c(R)}{Z_1^2}
\end{equation}
where $Z_1$ is the partition sum of a single polymer. Equivalently, $Z_1^2$ may be imagined as the partiton sum of a two-chain system whose centers of mass are infinitely far apart.

\subsection{Monte-Carlo algorithm}
The generation of ring polymers is accomplished using a well-tested lattice Monte Carlo algorithm, the bond fluctuation model~\cite{Carmesin1988, Deutsch1991}. Simulations are set up on a cubic lattice, where each monomer of the ring resides in the center of a unit cube and occupies all its eight vertices. Each monomer is connected by a bond to its two neighbors along the contour of the polymer, ensuring the connectivity of the chain. 
The lattice constant is taken to be unity and all measures of length in the following are given in multiples of the lattice constant. The length of the bonds between two neighboring monomers are allowed to fluctuate, but constrained to a set $\mathcal{B}$ of 108 vectors~\cite{Deutsch1991}. One Monte Carlo trial move encompasses randomly selecting one monomer and proposing a move to one of its nearest neighbors on the cubic lattice. Using these local moves combined with the constrained set of bond vectors $\mathcal{B}$, it has been shown that no bond-crossings can occur during the simulation run~\cite{Deutsch1991}, thus the topological state of the rings is preserved while propagating the system. Excluded volume interactions are taken into account by preventing a lattice site to be occupied by more than one monomer. 

Isolated ring polymers with chain lengths between $N=32$ and $N=2048$ are simulated. As only local moves are applied, consecutive conformations are highly correlated. To determine the relaxation time of a ring polymer, we evaluated the autocorrelation function $C(t)$~\cite{Binder2002} of the squared radius of gyration $A(t)=R_{g}^2(t)$ and from $C(t)$ we estimated the integrated autocorrelation time $\tau$ using Sokal's windowing procedure~\cite{Sokal1996, Bohn2009c}. Two subsequent conformations are considered independent after $5\tau$ Monte Carlo steps (MCS). In the following we define one MCS as $N$ trial moves, where $N$ is the chain length and therefore the total number of monomers in the system. Thus, in one MCS each monomer is proposed on average one move. For each chain length at least 10\,000 independent conformations were produced.

Simulations are performed on a lattice with periodic boundary conditions, but the algorithm keeps track of unfolded coordinates, such that the entire configuration space can be explored. The linear lattice size $L$ is chosen large enough, i.e. $L \gg \left<R_g^2\right>^{1/2}$, so that monomers do not come into contact in the lattice space by periodic backfolding of coordinates, otherwise biased conformations would result.  Lattice sizes used vary between $L=256$ up to $L=800$, depending on chain length $N$. 

\subsection{Calculation of the effective potential}\label{sec:methods:effpot}
\begin{figure}
 \includegraphics[width=\hsize]{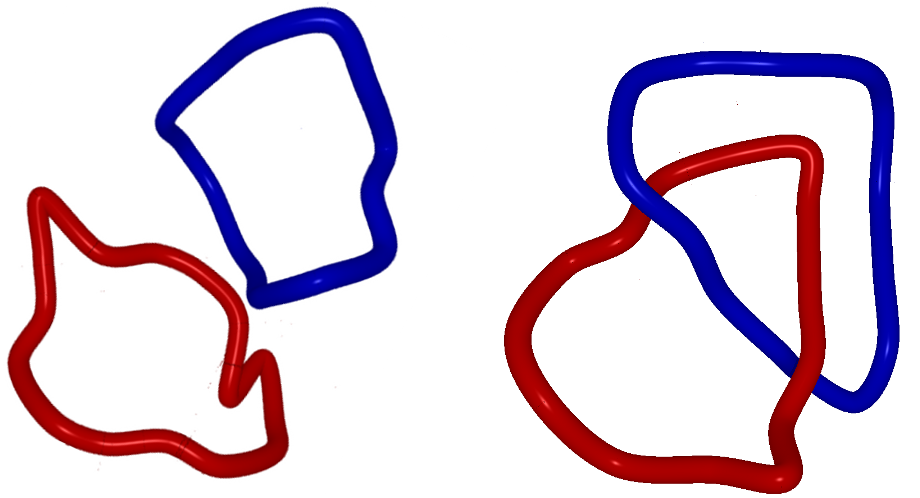}
 \caption{\label{fig:ring:illustration} Illustration of different topological states of two ring polymers. The left figure shows a non-catenated conformation with Gauss linking number $\Phi=0$. On the right, a catenated conformation is displayed having a linking number of $\Phi=1$.}
\end{figure}

Single ring conformations are used to calculate the average interaction energy between two polymers by means of equation~\eqref{eq:effpot:simple}. This procedure is well-established and has been applied in several studies~\cite{Dautenhahn1994, Hsu2004}. At first, the set of sampled conformations $\mathcal{C}$ is split into two parts of equal size. Then, we randomly pick two conformations, one from each subset. These two ring conformations are then positioned such that their centers of mass have a certain distance $r$, the angular positioning in doing so is chosen randomly. In order to ensure all monomers residing on lattice sites after shifting, a maximum distance inaccuracy of $\delta r \approx 0.87$ lattice units is accepted. In the next step, the interaction energy $E_i(r)$ of the two-ring conformation and the corresponding statistical weight $W_i(r) = \exp(-E_i / k_BT)$ are calculated. 

Repeating this operation for a number $K$ of randomly selected pairs of conformations, the effective potential at a center-of-mass distance $r$ is then given by
\[ U_{\mathrm{eff}}(r) = -k_BT \ln \frac{ \sum_{i=1}^K W_i(r) }{K} \]
The standard error of the effective potential is determined by randomly subdividing the $K$ two-chain conformations into $M=50$ smaller subsets and calculating the effective potential $U_m(r)$ for these subsets~\cite{Dautenhahn1994}. The standard error is then calculated by
\[ \Delta U_{\mathrm{eff}}(r) = \sqrt{\frac{1}{M} \sum_{m=1}^M (U_m(r)  - U_{\mathrm{eff}}(r))^2} \]

In a first step, only taking into account excluded volume interactions, the statistical weight $W_i(r)$ of a two-chain conformation is set to $W_i=0$ in case the excluded volume condition is violated, i.e. two monomers occupy the same lattice site, and $W_i=1$ in case it is not. 

However, care has to be taken using the bond fluctuation model. During the simulation run it is ensured that no bond crossings can occur; this, however, is not guaranteed if we just shift two conformations inside each other. There are two possible combinations of bond vectors, where the excluded volume condition is not violated, but where the bonds come into contact. Consider the two bonds spanned by the following four monomers $\{ (0,0,0) \rightarrow (3,1,0), (1,2,0)\rightarrow(2,-1,0)\}$. Both bond vectors $(3,1,0)\in\mathcal{B}$ and $(1, -3, 0)\in \mathcal{B}$ are valid and the excluded volume condition is satisfied. The second problematic case is $\{ (0,0,1) \rightarrow(2,2,0), (0,2,0)\rightarrow(2,0,1)\}$.  As can be easily seen, the participating beads are stuck in these positions, there is no valid move for any of the beads using the restrictions of bond vectors the algorithm is subjected to. Therefore, during a simulation run such a situation can never happen, assuming the start configuration is chosen properly. Here, we test each two-chain conformation, after positioning them with their centers of mass a given distance $r$ apart, on such a situation. If a bond crossing occurs, a weight of $W_i=0$ is assigned to this conformation. 

In this study, we are furthermore interested in the topological state of a two-ring conformation. We calculate the weight factor differently whether we consider a preserved topological state or not. In the case that we force a certain topological state, we have to add another criterion for accepting a two-chain conformation (i.e. assigning a weight $W_i \neq 0$). Two rings that are positioned at a center-of-mass distance $r$ might be non-catenated or catenated. Furthermore, the degree of catenation might vary (see illustrations in Fig.~\ref{fig:ring:illustration}). We determine the topological state by a topological invariant: the ``Gauss linking number''~\cite{Otto2004}. This invariant has been used in analytical studies on interlinked rings to keep track of a certain topology~\cite{Otto2004, Otto1998}. It is defined by
\begin{equation}\label{eq:gln}
 \Phi(C_1, C_2) = \frac{1}{4\pi} \oint_{C_1} \oint_{C_2}  \frac{\left<d\mathbf{r}_1 \times d\mathbf{r}_2, \mathbf{r}_1 - \mathbf{r}_2\right> }{|\mathbf{r}_1 - \mathbf{r}_2|^3} 
\end{equation}
The closed line integrals are evaluated along the contours of the two rings, denoted by $C_k\, (k=1,2)$. The vector function $\mathbf{r}_k=\mathbf{r}_k(s)$ denotes the three-dimensional coordinates of the ring polymers, parameterized e.g. by the contour length $s$. 

The Gauss link invariant has the advantage that it is easy to evaluate, although it has the disadvantage that it is not one-to-one. The  ``Whitehead link''~\cite{Otto2004} is one example for a two-ring conformation having the same linking number as non-catenated rings. However, for this link, self-intersections of one ring are necessary, which do not arise during the simulation runs. Therefore we consider the Gauss linking number as an excellent quantity to determine the topological state. 
The result of the integral is an integer for non-overlapping chains, especially $\Phi=0$  for non-catenated chains and  $\Phi=1$ for simple catenated chains (Fig.~\ref{fig:ring:illustration}). We numerically evaluate this integral to classify the topological state of a two-ring conformation and assign a weight $W_i=0$ for all two-ring conformation not in the topological state we are interested in.  Thereby we are able to calculate the effective potential separately for any given topological state.

\section{Results}\label{sec:results}
\subsection{The effective potential of self-avoiding walks}\label{sec:results:saw}
\begin{figure*}
 \includegraphics[width=\hsize]{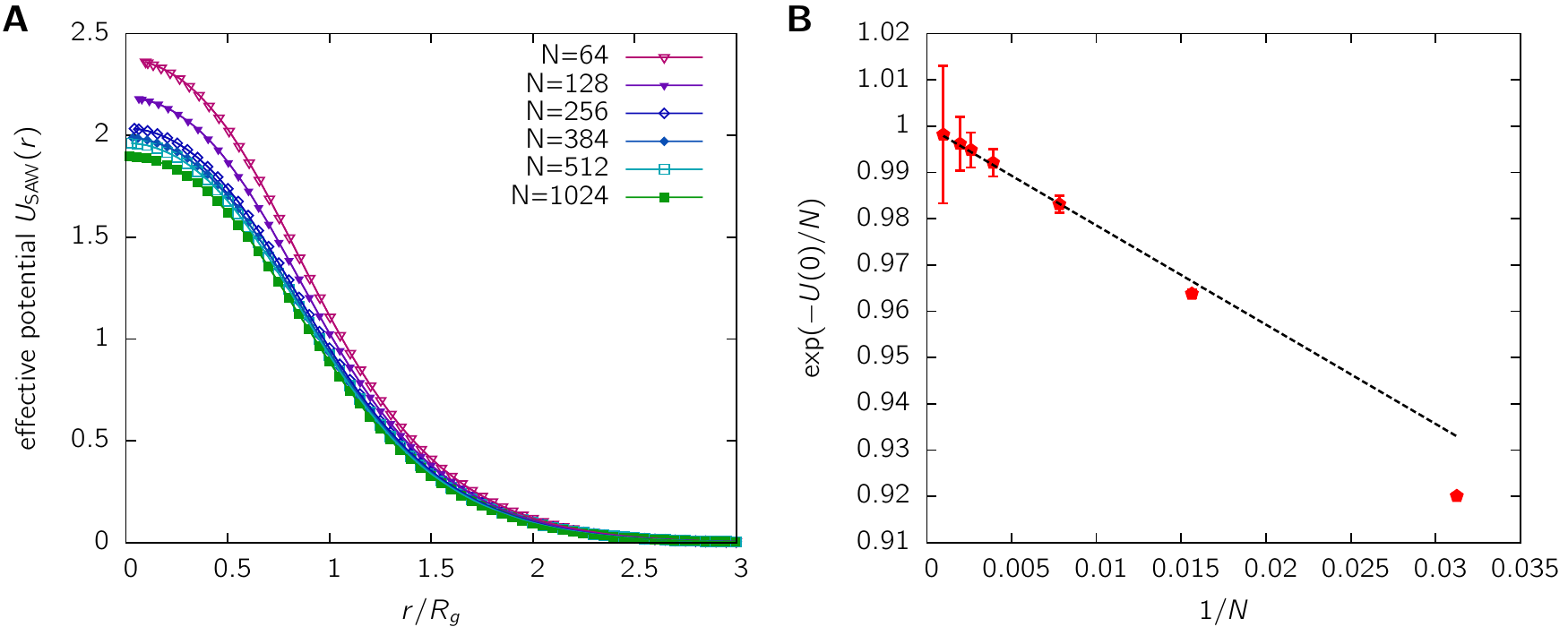}
 \caption{\label{fig:effpot:saw} Effective potential acting between the centers of mass of two self-avoiding walks. $\mathbf{A.}$ This panel displays the effective potential $U_{\text{SAW}}(r)$ for different chain lengths. The $x$-axis is scaled by the root mean squared radius of gyration $R_g$ of isolated chains. Errors are smaller than the point size and therefore not shown. The effective potential is dependent on chain length. $\mathbf{B.}$ Scaling plot of the effective potential at full overlap. The effective potential $U_{\text{SAW}}(0)$ reaches a finite value for large chain length as expected from scaling theory (cf. equation~\eqref{eq:effpot:r0}).}
\end{figure*}

We first recapitulate the results for the effective potential between the centers of mass of two self-avoiding walks both to validate the algorithm used and to relate the strength and form of the potential for linear polymers to their ring counterparts.

A simple mean field argument can be devised for the entropic repulsion of two polymer coils with excluded volume at full overlap, i.e. with zero center-of-mass separation. It has been first proposed by Flory and Krigbaum~\cite{Flory1950} and later been corrected in other studies~\cite{Daoud1975, Grosberg1982}. The overlap volume is given by  $V\sim R_g^3$ where $R_g$ is the radius of gyration of one polymer coil. A mean-field argument yields the effective potential at full overlap~\cite{Grosberg1982}: Scaling arguments~\cite{Daoud1975}, which take into account the correlation hole effect~\cite{Gennes1979}, show that the probability $p_1$ of one monomer of chain $A$ being in contact with any monomer of $B$ is given by $p_1 = (n a^3) ^ {\frac{1}{3\nu-1}}$. Thus the probability that none of the monomers are overlapping is given by  $p=(1-p_1)^N = (1 - \frac{b'}{N})^N$. The effective potential can then be calculated as the logarithm of this probability, 
\begin{equation}\label{eq:effpot:r0}
  U(r=0) = -N \ln \left( 1 - \frac{b'}{N} \right) 
\end{equation}
In the limit of infinite chain length $N\rightarrow\infty$, the effective potential at full overlap approaches a constant value in the order of $k_BT$. Consequently, linear polymer coils  with excluded volume have a rather soft potential, allowing for a high level of mutual penetration.

Results for the self-avoiding walk are displayed in Fig.~\ref{fig:effpot:saw}. The effective potential $U_{\text{SAW}}(r)$ in Fig.~\ref{fig:effpot:saw}A reveals a decrease with growing chain length, inconsistent with the mean-field argument by Flory~\cite{Flory1950} but consistent with eq.~\eqref{eq:effpot:r0}. The effective potential at full overlap is shown in a scaling plot (cf. Ref.~\cite{Bolhuis2001}) in Fig.~\ref{fig:effpot:saw}B. Data points are on a straight line as expected, deviations are most likely due to lattice effects: When the two polymer coils are brought to a certain distance, the nearest lattice site is chosen, such that the distance can deviate from the target distance $r=0$. These deviations are more pronounced the smaller $N$. 

\subsection{Rings have a stronger repulsion than linear polymers}\label{sec:results:rings}
\begin{figure*}
 \includegraphics[width=\hsize]{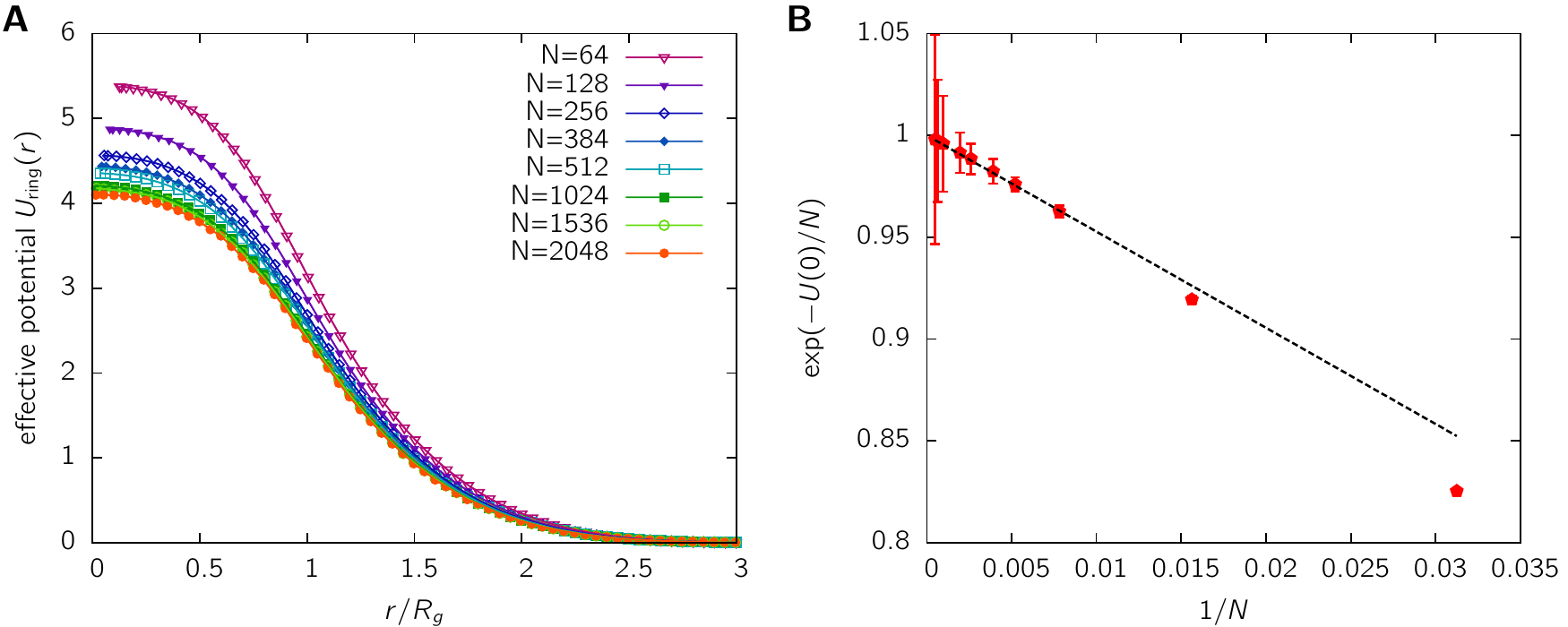}
 \caption{\label{fig:effpot:nt} Effective potential $U_{\text{ring}}(r)$ of ring polymers disregarding their mutual topological state. \textbf{A.} This panel displays the effective potential in units of $k_BT$ for several chain lengths. Errorbars are not shown as they are smaller than the point size. For reasons of comparison, the $x$-axis is scaled with the radius of gyration of the corresponding isolated ring polymer. \textbf{B.} The scaling law for the effective potential at full overlap follows the theory from eq.~\eqref{eq:effpot:r0} allowing for a linear extrapolation to the infinite chain limit when plotting $e^{-U(0)/N}$ vs $1/N$. For the extrapolation, only chain lengths larger than $N=256$ are used.}
\end{figure*}

\begin{figure}
  \includegraphics[width=\hsize]{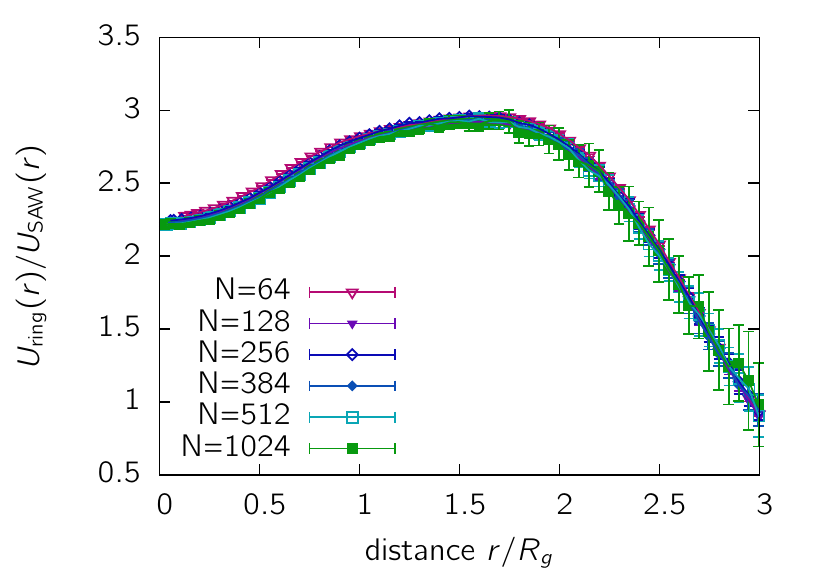}
  \caption{\label{fig:effpot:ntsaw} Comparison of the effective potentials for ring polymers and self-avoiding walks. The ratio $U_{\text{ring}}(r)/U_{\text{SAW}}(r)$ is plotted against the center-of-mass distance $r$. The $x$-axis is scaled by the root mean squared radius of gyration $R_g$ of isolated chains for reasons of comparison.}
\end{figure}

To study the  effective potential between ring polymers, we first neglect topological constraints between the two rings when bringing them in proximity. However, each single ring obeys the topological constraint of unknottedness by virtue of the simulational method. Thus the effective potential is given by the logarithm of the probability that the ring conformations do not occupy at least one lattice site in common. 

Figure~\ref{fig:effpot:nt}A shows the effective potential $U_{\mathrm{ring}}(r)$ for ring polymers where mutual topological constraints are neglected, i.e. the linking number between the two ring polymers is arbitrary. The effective potential again is dependent on chain length $N$, however, differences become smaller for large chain lengths. The mean field argument for the effective potential at full overlap established in eq.~\eqref{eq:effpot:r0} does not depend on the connectivity of the chain. Indeed, it is not only valid for linear polymers, but also for rings. Figure~\ref{fig:effpot:nt}B shows the effective potential at full overlap $U_{\mathrm{ring}}(0)$ such that a linear extrapolation to $N\rightarrow\infty$ becomes possible. Clearly, the effective potential adopts a constant value in this limit. 

Interestingly, the repulsive interactions between two ring polymers are much stronger than for their linear counterparts of equal chain length $N$. The ratio  $U_{\mathrm{ring}}(r)/U_{\text{SAW}}(r)$ is displayed in Fig.~\ref{fig:effpot:ntsaw} and has a value of about $3$ for center of mass (CM) separations below $2 R_g$.
%

\subsection{Topological constraints induce additional repulsion}\label{sec:results:top}
\begin{figure}
\includegraphics[width=\hsize]{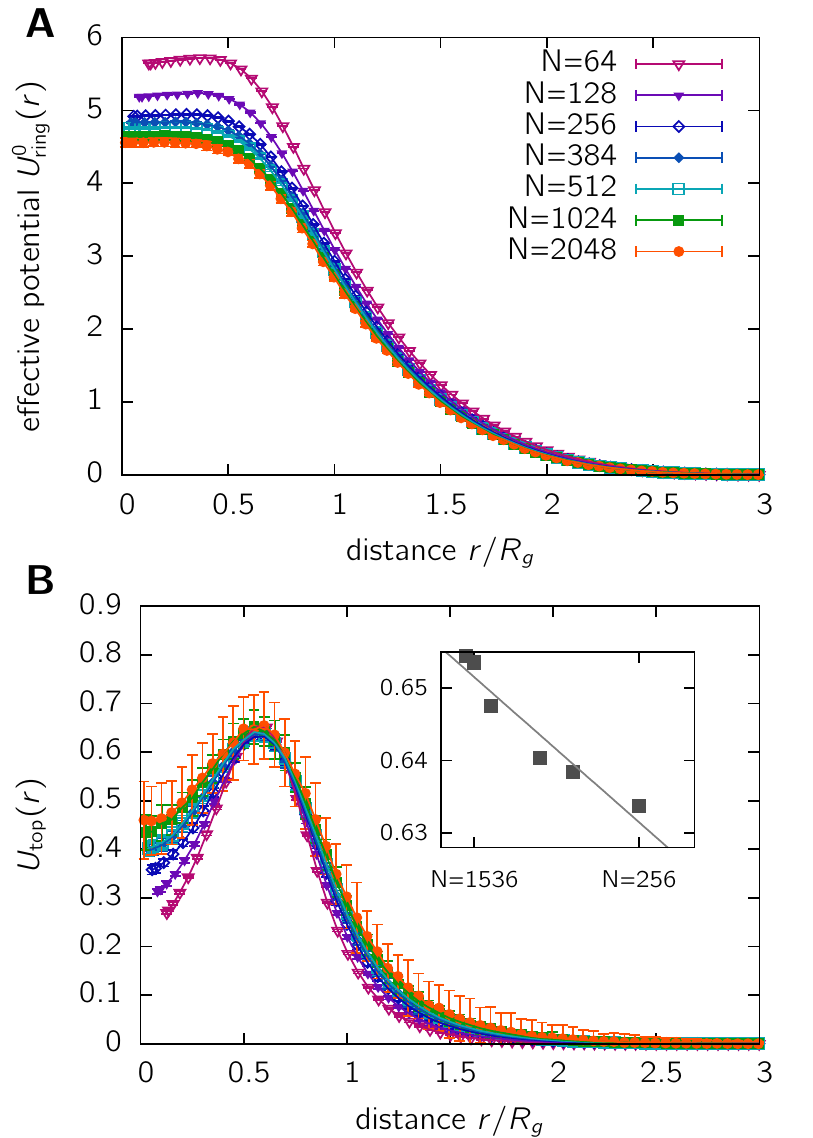}
\caption{\label{fig:effpot:nc} \textbf{A. } The effective potential $U^0_{\text{ring}}(r)$ for non-catenated rings. \textbf{B. } The topological potential $U_{\text{top}}(r)$ for non-catenated rings is obtained as the difference between $U_{\text{ring}}^0(r)$ and $U_{\text{ring}}(r)$. It displays the potential arising purely from the topological non-catenation constraint. The inset shows the maximum of the topological potential depending on the chain length $N$. The maxima are obtained by a fit of $U_{\text{ring}}(r)$ to a third-order polynomial in the interval $[0.3:0.7]$.}
\end{figure}

In the last section we have shown that the repulsive forces acting between two ring polymers whose centers of mass are brought in close proximity are much stronger than for linear self-avoiding walks, the potential being about $3$ times larger. This rise in strength results from the increased density of monomers induced by the ring structure. Here we focus on the additional effect topological constraints induce on the potential of mean force between the polymers. Much attention has been paid to the influence of topological constraints on ring polymers. The interaction between non-linked ring polymers without excluded volume has been investigated for chains of length up to $N=80$~\cite{Frank-Kamenetskii1975}. In a recent study~\cite{Bohn2009c} we have analyzed the behavior of two catenated rings. It turned out that while the dimensions of catenated rings show the same scaling behavior than their linear counterparts, the shape changes significantly. The topological states of circular DNA have been analyzed in various publications~\cite{Podtelezhnikov1999, Rybenkov1997}. The studies on non-catenated and unknotted ring polymers in the melt revealed that their dimensions change dramatically~\cite{Mueller1996, Mueller2000, Vettorel2009}: They behave like compact polymers with a radius of gyration scaling with $N^{1/3}$. The dominant entropic force driving this compaction stems from the topological constraint of non-catenation. To our knowledge, a quantitative study of this topological potential has not been considered yet.  
In the framework of effective interactions, the topological non-catenation constraint can be included easily. As explained in the section~\ref{sec:methods:effpot}, we determine the topological state of a two-ring conformation by means of its Gauss linking number $\Phi$ (eq.~\eqref{eq:gln}). A Gauss linking number of zero indicates the non-catenation of the two ring polymers. To obtain the potential of mean force under the non-catenation constraint, we calculate the fraction $p$ of two-ring conformations with both zero linking number and fulfilled excluded volume constraint. The effective potential is then given by $U_{\text{ring}}^0(r) = -\ln p$ (the index $0$ indicates the Gauss linking number $\Phi=0$). The resulting  potential $U_{\text{ring}}^0(r)$ is shown in Fig.~\ref{fig:effpot:nc}A. We find pronounced deviations from a Gaussian shape for small CM separations $r \lesssim 1 R_g$. 

From the effective potential $U_{\text{ring}}^0(r)$ of non-catenated rings and $U_{\text{ring}}(r)$ of rings without topological constraints, the topological potential $U_{\text{top}}(r)$, i.e. the part of $U_{\text{ring}}^0(r)$ stemming purely from the non-catenation constraint, can be obtained,
\[ U_{\text{top}}(r) = U_{\text{ring}}^0(r) - U_{\text{ring}}(r) \]
The topological potential is shown in Fig.~\ref{fig:effpot:nc}B. It has a maximum at about $r/R_g\approx 0.55$, a value which basically does not change with chain length $N$. For larger CM separations $U_{\text{top}}(r)$ drops to zero, as less and less two-ring conformations will be catenated when increasing $r$. There is also a decrease in the topological potential when going to small CM separations $r < 0.5 R_g$. This decrease becomes smaller for larger chain length $N$. The reason for this behavior is that for small $r$ (and small $N$), two-ring conformations not satisfying the non-catenation constraint, often do not satisfy the excluded volume constraint either, such that the conformations are not counted for the purely topological interaction $U_{\text{top}}$. 
However, more interesting  is the maximum repulsion exerted by the non-catenation constraint. We find that the topological potential is asymptotically $U_{\text{top}}^{\mathrm{max}}  \approx 0.6 - 0.7 \: k_BT$  for $(N\rightarrow \infty)$
at $r/R_g \approx 0.55$ (see inset of Fig.~\ref{fig:effpot:nc}B). In numbers, this means that only about $48\%$ ($\exp(-0.65) \approx 0.52$) of all two-ring conformations with fixed center-of-mass distance which would be allowed with respect to excluded volume are actually allowed with respect to topological constraints. In other words, only half of the two-ring conformations satisfying the excluded volume constraint are actually accessible when the non-catenation constraint is considered. Thus, the non-catenation constraint decreases the number of accessible conformations at small CM separations considerably. 


Interestingly, the topological interaction is on the order of magnitude of $1\,k_BT$. This value was used for a crude estimate of the free energy of rings in a melt by Cates and Deutsch~\cite{Cates1986}, where they assumed that the free energy cost for a contact between two rings is about $1\,k_BT$, hence in a three-dimensional melt the topological contribution to the free energy is $F\approx k_BT R^3 / N$. Although highly simplistic, this scaling exponent $\nu=2/5$ was later reproduced by simulational results~\cite{Mueller1996}; more recent results, however, suggest that this behavior is only a cross-over to the behavior of compact lattice animals~\cite{Mueller2000}.

\subsection{Dimensions of two-ring conformations}\label{sec:results:dimensions}
\begin{figure}
\includegraphics[width=\hsize]{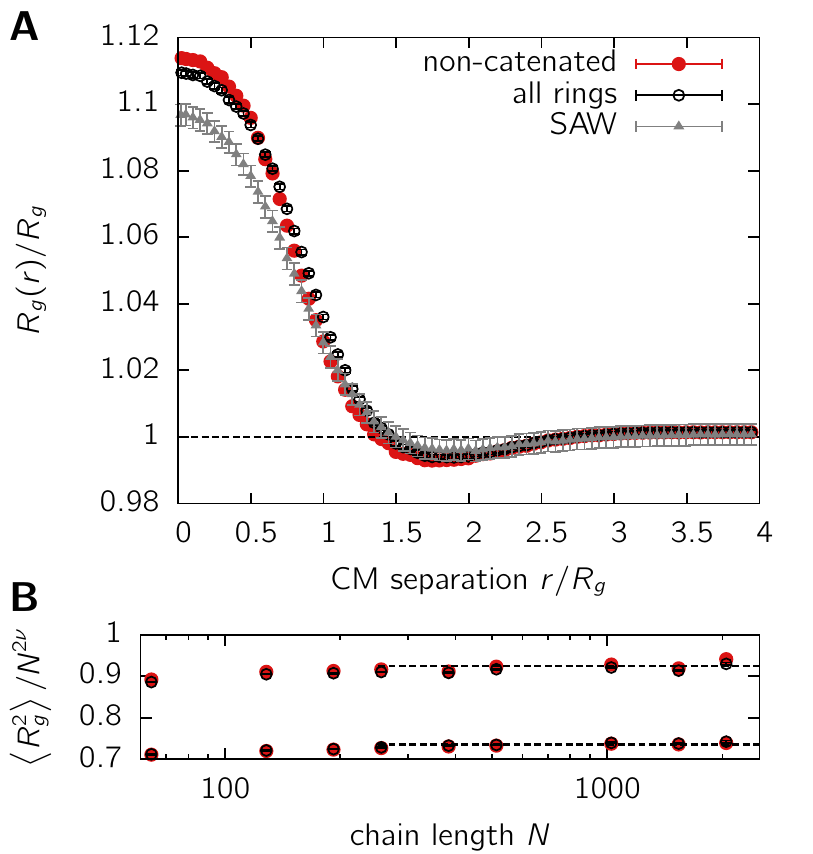}
\caption{\label{fig:dim:rgyr} \textbf{A. }Mean squared radius of gyration for self-avoiding walks (SAW), ring polymers without mutual topological interactions  and non-catenated rings  vs. distance between the centers of mass. Radii of gyration are scaled by the corresponding values $R_g$ of the isolated chains. 
\textbf{B. } Scaling of the radius of gyration with chain length. The leading term $N^{2\nu}$ of a self-avoiding walk ($\nu=0.588$) is divided out to see deviations from the linear chain behavior. }
\end{figure}
\begin{figure}
\includegraphics[width=\hsize]{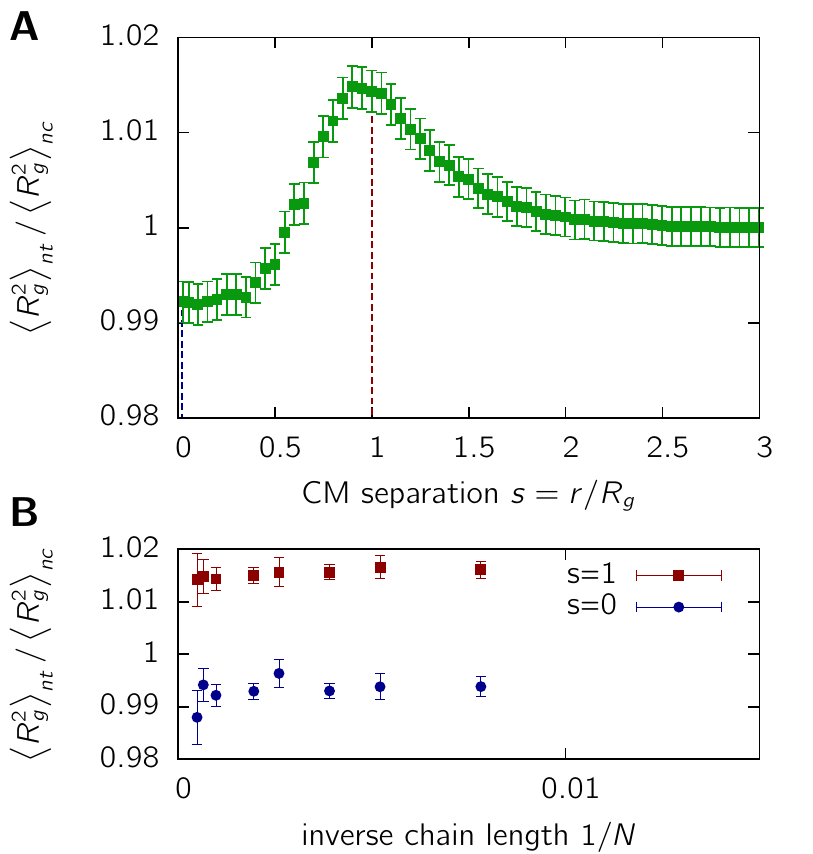}
\caption{\label{fig:dim:rgyr:ratio} \textbf{A. } Ratio between the mean squared radii of gyration of rings without mutual topological interactions (NT) and non-catenated rings (NC) vs. scaled distance $r/R_g$ between the centers of mass. Results are shown for a chain of length $N=1024$. \textbf{B. } The ratio is plotted against the inverse chain length allowing for an extrapolation to the infinite chain limit $N\rightarrow\infty$. Data is shown for two CM separations: $s=0$ and $s=1$ to analyze whether deviations remain in the limit of large chain length. }
\end{figure}
\begin{figure}
\includegraphics[width=\hsize]{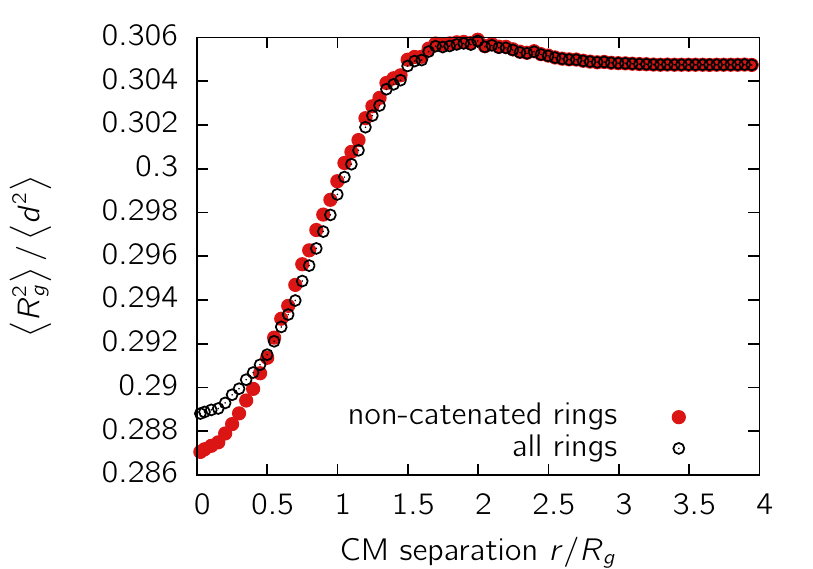}
\caption{\label{fig:dim:rgyr:d:ratio} Ratio between the mean squared radius of gyration and mean squared ring diameter vs center-of-mass separation $r$. Data is shown for a chain length of $N=1024$. Results are shown for two-ring conformations without mutual topological interactions (open circles) and non-catenated ring conformations (solid red circles).}
\end{figure}

When two polymers are brought closer and closer together, lowering the distance between the centers of mass, not only the repulsive interaction increases dramatically. Also the conformational properties of the polymers are affected~\cite{Dautenhahn1994} in an effort of minimizing the free energy. Here we investigate how the dimensions of a two-ring conformation change compared to a self-avoiding walk when approaching isolated polymers. 

There are two measures of the dimensions of the chain, which, however, exhibit the same scaling: the mean squared radius of gyration $\left<R_g^2\right>$ and the mean squared ring diameter $\left<d^2\right>$ (see Ref.~\cite{Bohn2009c}). Figure~\ref{fig:dim:rgyr}A shows the root mean squared radius of gyration $R_g(r)=\sqrt{\left<R_g(r)^2\right>}$ for self-avoiding walks (SAW), ring polymers without mutual topological interactions (arbitrary linking number  $\Phi$) and non-catenated rings (linking number $\Phi=0$) in dependence of the scaled center-of-mass separation $s=r/R_g$. The data are scaled with the radius of gyration for isolated polymers $R_g=R_g(r=\infty)$ (i.e. the radius of gyration of a polymer at infinite distance away from the second one) for reasons of comparison. For small CM separations ($r < R_g$) the radius of gyration strongly increases and its dimensions are more than 10\% larger than in the isolated case. This is consistent with the results for self-avoiding walks by Dautenhahn and Hall~\cite{Dautenhahn1994}. The increase in dimension at separations $r \ll R_g$ originates from the need of the chains to create space for the monomers of the second chain. 
At intermediate separations of about $r/R_g \approx 1.5-2$, the radius of gyration attains a smaller value, thus the polymer effectively undergoes compaction. 
Both the compaction at intermediate separations as well as the swelling at high overlap remain in the limit of infinite chain length. An extrapolation of the radius of gyration at full overlap ($s=0$) to $N\rightarrow\infty$ shows that $R_g(r=0)/R_g\rightarrow 1.109(1)$ for rings without topological interactions and $R_g(r=0)/R_g\rightarrow 1.111(1)$ for non-catenated rings (see Supplementary Figure 1~\cite{JCPringSI1}). 


One might ask, whether this change in dimensionality is accompanied by a change in the scaling law. We know from several studies that isolated ring polymers follow a similar scaling law as linear polymers $\left<R_g^2\right> \sim N^{2\nu}$ where $\nu\approx 0.59$~\cite{Sikorski1994, Brown2001, Bohn2009c}. However, non-catenated rings in a melt behave like compact polymers with a scaling exponent $\nu=1/3$~\cite{Mueller2000, Vettorel2009}.  Figure~\ref{fig:dim:rgyr}B shows the scaling of the radius of gyration with chain length in a log-log plot for two center-of-mass separations: $r=0$ (full overlap, region where the chain is swollen) and $r=1.7R_g$ (region where the chain is compacted). The leading order term $N^{2\nu}$ of the self-avoiding walk behavior ($\nu=0.588$~\cite{Le1980}) is divided out. The data adopt a constant value in the limit of large chain length $N$ independent of CM separation, indicating that the scaling behavior is independent of separation. As this also applies to rings with the non-catenation constraint, the compaction found in a polymer melt of non-catenated rings seems to require several chains, exerting forces from all sides on the polymer coil.

Our results show that differences between the radius of gyration $\left<R_g^2\right>_{nt}$ of rings without topological constraints and $\left<R_g^2\right>_{nc}$ of rings obeying non-catenation are rather subtle. The ratio of the radii of gyration $\left<R_g^2\right>_{nt}/\left<R_g^2\right>_{nc}$  with respect to the CM separation is shown in Fig.~\ref{fig:dim:rgyr:ratio}A for a chain length of $N=1024$. For small separations $r < 0.5 R_g$ non-catenated rings are smaller, while for intermediate separations ($r \sim 1 R_g$) they are larger. The dependence of these results on the chain length $N$ is given in Fig.~\ref{fig:dim:rgyr:ratio}B for two different CM separations. The differences are asymptotically stable both in the regime of small as well as intermediate separations. At full overlap ($s=0$), non-catenated rings are larger in size by about 0.8\% in the asymptotic limit $N\rightarrow \infty$, for intermediate separations ($s=1$), rings without topological constraints are larger by about 1.5\%. Similar results are found for the mean square ring diameter (data not shown).

As isolated ring polymers as well as self-avoiding walks only have one length scale, different measures of dimension should yield a constant ratio. For a random walk, the ratio between the mean squared radius of gyration and the end-to-end distances is $1/6$. For isolated ring polymers, the ratio $\left<R_g^2\right>/\left<d^2\right>$ extrapolates to $0.3053(2)$ in the asymptotic limit, while catenated rings display a ratio of $0.2995(3)$~\cite{Sikorski1994, Bohn2009c}. For the case studied here, the ratio between the radius of gyration and ring diameter $\left<R_g^2\right>/\left<d^2\right>$ changes significantly for different CM separations (see Fig.~\ref{fig:dim:rgyr:d:ratio}). From its isolated chain value the ratio decreases  when bringing the centers of mass closer together. Extrapolation to infinite chain length $N\rightarrow\infty$ shows that the differences remain even in this limit (Supplementary Figure 2~\cite{JCPringSI2}).  At full overlap ($s=0$) the ratio extrapolates  to $\left<R_g^2\right>/\left<d^2\right>=0.2874(3)$ for non-catenated rings and $\left<R_g^2\right>/\left<d^2\right>=0.2891(3)$ for rings without topological interactions.

\subsection{Shape of two-ring conformations}\label{sec:results:shape}
\begin{figure*}
\includegraphics[width=\hsize]{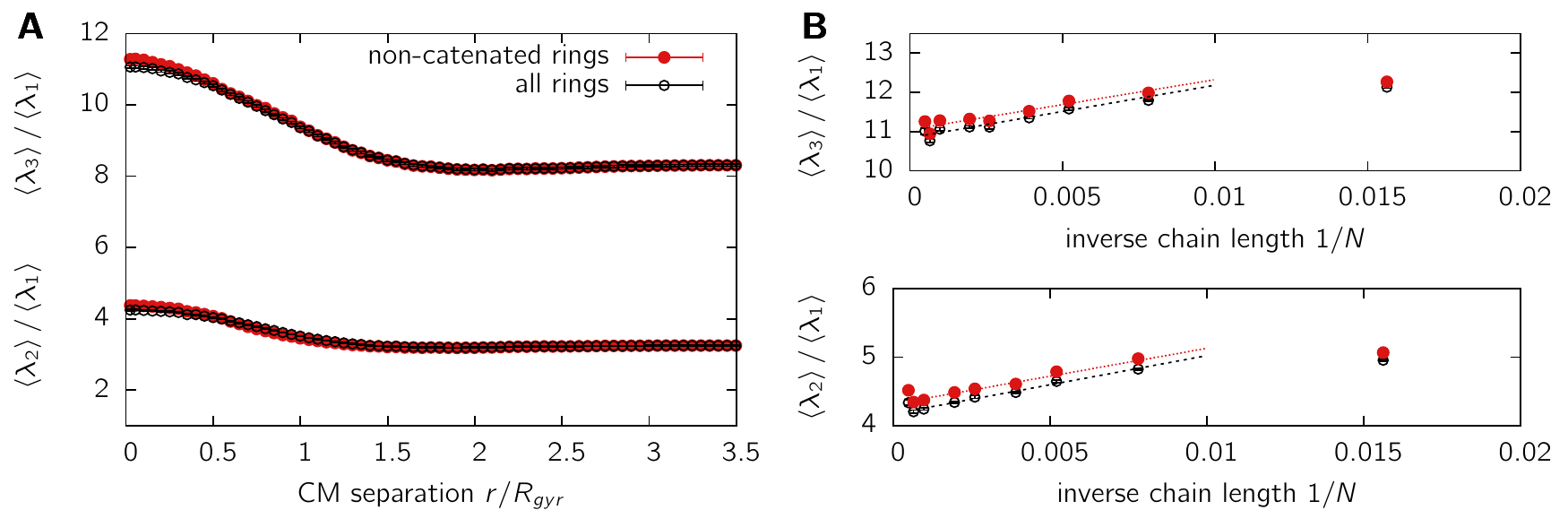}
\caption{\label{fig:shape:ratios} \textbf{A.} Ratios between the gyration tensor's eigenvalues $\lambda_1 \leq \lambda_2 \leq \lambda_3$ in relation to the center of mass distance $s=r/R_g$. Data is shown for a system with chain length $N=1024$ for two-ring conformations without mutual topological interactions (open circles) and non-catenated ring conformations (solid red circles). \textbf{B.} Finite-size behavior of the ratios $\left<\lambda_3\right>/\left<\lambda_1\right>$ and $\left<\lambda_2\right>/\left<\lambda_1\right>$ for a center-of-mass separation of $s=0$.}
\end{figure*}
A typical measure of the shape of a polymer coil is its gyration tensor~\cite{Bruns1992, Rawdon2008}. It is defined by 
\begin{equation}
 S_{mn} = \frac{1}{N} \sum_{i=1}^N r_m^{(i)} r_n^{(i)}
\end{equation}
Here $\mathbf{r}^{(i)}$ is the coordinate vector of the $i^{\text{th}}$ monomer and the subindex denotes its cartesian components. 
The matrix $\mathbf{S}$ is symmetric and positive semi-definite, thus it can be transformed to a diagonal matrix where the three eigenvalues $\lambda_1 \le \lambda_2 \le \lambda_3$ give the squared lengths of the principal axes of gyration of the associated gyration ellipsoid. In fact, the sum of the eigenvalues gives the squared radius of gyration $R_g^2 = \lambda_1 + \lambda_2 + \lambda_3$. 

In a recent study~\cite{Bohn2009c} on catenated rings we have found that the shape of catenated rings differs significantly from the shape of isolated ring polymers. This shows up in the ratios of the averaged eigenvalues  $\left<\lambda_3\right> : \left<\lambda_2\right>: \left<\lambda_1\right>$, which are obtained from extrapolation to the asymptotic limit as $8.23(2):3.22(2):1$ for isolated rings and $8.87(3):3.56(2):1$ for catenated rings.

We investigate how the shape of rings changes when approaching the centers of mass, both for rings without topological interactions as well as for conformations, which obey the non-catenation constraint. Figure~\ref{fig:shape:ratios}A displays the results for a chain of length $N=1024$. The polymers get strongly elongated for small center-of-mass separations $s=r/R_g$. This elongation at full overlap $s=0$ remains in the limit of large chain length $N\rightarrow \infty$ as is displayed in Fig.~\ref{fig:shape:ratios}B. For non-catenated rings, the effect of elongation is slightly more pronounced than for rings, where topological effects are neglected.
At full overlap we find for non-catenated rings that the ratios $\left<\lambda_3\right> : \left<\lambda_2\right>: \left<\lambda_1\right>$ are asymptotically $11.05(8) : 4.32(3) : 1$, while for rings without mutual topological interactions we have $10.84(6) : 4.17(2) : 1$.

While the analysis of the radius of gyration indicated that the spatial extend of rings becomes larger, we have found here that this opening up does not lead to a complete mixing of the two rings, which would result in a more spherical structure; rather rings elongate strongly in one direction.

\subsection{Orientation and Alignment of rings}\label{sec:results:align}
\begin{figure}
\includegraphics[width=\hsize]{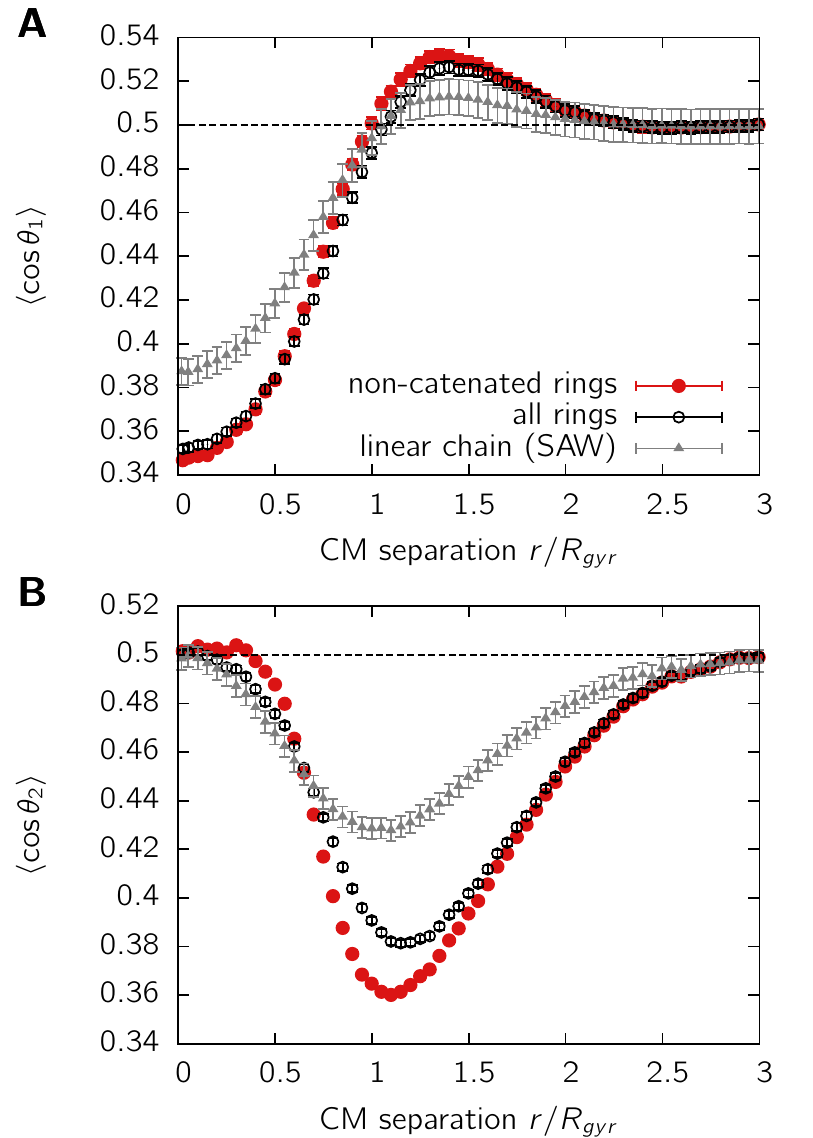}
\caption{\label{fig:alignment:cos} \textbf{A.} Average angle $\left<\cos\theta_1\right>$ between the largest principal axes of the gyration ellipsoids of a two-chain conformation in relation to its center of mass separations. The black line corresponds to a random orientation of the two axes. Results are shown for self-avoiding walks (grey triangles), non-catenated rings (solid red circles) and rings without mutual interactions (open black circles). \textbf{B. } Average angle $\left<\cos\theta_2\right>$  between the vector connecting the centers of mass and the largest principal axis of one polymer coil in relation of the CM separation. }
\end{figure}	
\begin{figure}
 \includegraphics[width=\hsize]{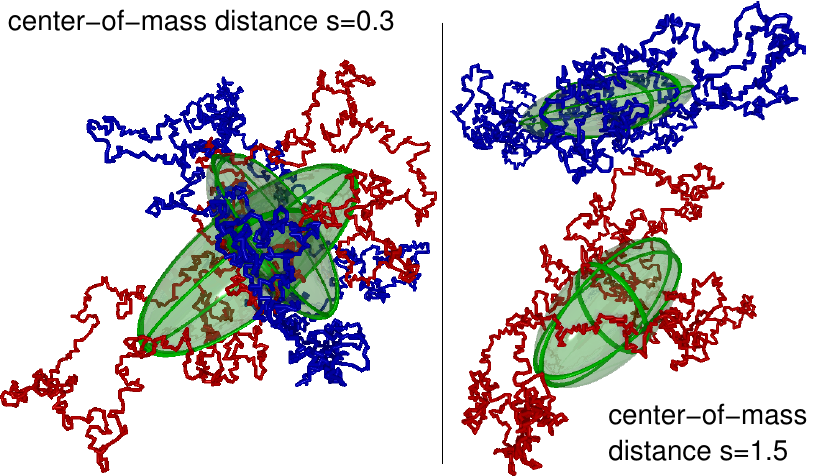}
 \caption{\label{fig:alignment:samples} Sample conformations of non-catenated rings forced to a certain center-of-mass distance. The left-hand figure displays a conformation of two non-catenated rings at nearly full overlap ($s=r/R_g=0.3$). The two rings markedly align perpendicularly. The right-hand figure shows a two-ring conformation of non-catenated rings at separation $s=r/R_g=1.5$ displaying a tendency for parallel alignment. }
\end{figure}

To investigate the interactions between the two rings at full overlap in more detail, we look at the mutual alignment and orientation. The alignment of the gyration ellipsoids of both rings with respect to each other can be investigated via two measures: Firstly, the average angle $\left<\cos\theta_1\right>$ between the largest main axes of the ellipsoids, secondly the average angle $\left<\cos\theta_2\right>$ between the vector connecting the centers of mass of both rings and the largest main axis of one ellipsoid. Its behavior has been observed for catenated rings~\cite{Bohn2009c}. It was found that the main axes of the ellipsoids tend to align perpendicular for short CM separations, while for large CM separations, the alignment becomes more and more parallel. 

The results for two-ring conformations without mutual topological constraints and non-catenated ring conformations as well as SAWs are shown in Fig.~\ref{fig:alignment:cos}. The black line at $\left<\cos\theta_i\right>=0.5$ corresponds to the average angle for a random orientation of the vectors (Note that the angle $\theta_i$ can adopt only values between zero and 90 degrees). The average angle $\left<\cos\theta_1\right>$ between the gyration tensors  main axes (cf. Fig.~\ref{fig:alignment:cos}A) displays a pronounced perpendicular orientation for small CM separations $r < R_g$. This effect is markedly stronger for ring polymers than for linear ones. At intermediate separations of $r\approx1-2R_g$, however, a regime is found where a slightly parallel alignment of the rings is preferred compared to a random orientation. For large center of mass separations, a random orientation is approached; in this regime, the rings are nearly independent and therefore do not influence each other. An extrapolation to the infinite chain length regime at full overlap ($s=0$) yields an average angle $\left<\cos\theta_1\right>$ of 0.3543(7) for rings without topological interactions, 0.3492(6) for non-catenated rings and 0.3897(2) for self-avoiding walks. Details of the extrapolation are shown in the Supplementary Figure 3~\cite{JCPringSI3}. 

The angle $\left<\cos\theta_2\right>$ (see Fig.~\ref{fig:alignment:cos}B) shows a nearly random alignment for CM separations of $r \approx 0$. For intermediate $r\approx1-2R_g$, the alignment becomes pronouncedly more perpendicular compared to a random alignment.  This effect is much stronger for non-catenated rings than for rings without mutual topological constraints. Clearly, ring polymers display a much stronger tendency to align non-randomly than self-avoiding walks. In the regime of perpendicular alignment at $s=1.2$ the asymptotic values of 
0.3836(5) (rings without topological interactions), 0.3661(4) (non-catenated rings) and 0.4327(3) (SAWs) are found for the average angle $\left<\cos\theta_2\right>$~\cite{JCPringSI3}.

A recurrent motif in the analysis of shape, dimensions and orientation is a change in structure from polymers at full overlap to polymers at intermediate separations ($r \approx 1 - 2 R_g$). While they are strongly elongated and aligned perpendicular at short separations, at intermediate separations  compaction and parallel alignment is observed. Clearly, this results from a tendency to minimize the overlap area between both rings. At full overlap, this is accomplished best by a strong elongation and perpendicular alignment (Fig.~\ref{fig:alignment:samples}A). When rings are separated further  apart, the gyration ellipsoids can avoid intermingling by aligning in parallel (Fig.~\ref{fig:alignment:samples}B). The restricted conformational space due to the presence of the other ring results in the observed compaction of the radius of gyration compared to isolated chains. 

Both the topological constraints and the ring connectivity of the chain have a strong influence on how the polymers are aligning when brought close together. In general, the view emerges that rings have a more aligned and ordered state than linear chains, an effect which is even amplified by the non-catenation constraint.


\section{Conclusions}
In this study we investigated the effect of the non-catenation constraint on a system with two ring polymers. 
Our main focus was on the quantitative analysis of the strength of the interactions between ring polymers, including their topological interactions. For this purpose, we sampled conformations of isolated rings using a well-established Monte Carlo method.  We evaluated the potential of mean force using the idea of effective interactions~\cite{Likos2001} following the method introduced by Dautenhahn and Hall~\cite{Dautenhahn1994}. The topological state of a two-ring conformation is analyzed by means of the Gauss linking number. 

We found that the effective potential at full overlap adopts a constant value both for linear chains and ring polymers (Figs.~\ref{fig:effpot:saw} and~\ref{fig:effpot:nt}). However, the repulsive interaction between the centers of mass is about 3 times larger for ring polymers  than the effective potential for corresponding linear chains (Fig.~\ref{fig:effpot:ntsaw}) at small center-of-mass separations $r$. 

If the ring polymers have to stay in the fixed mutual topological state of non-catenation, the space of accessible conformations is further reduced, thus the effective potential increases. We have evaluated the strength of these interactions -- the topological potential $U_{\text{top}}(r)$ -- resulting purely from topological constraints. We find that the non-catenation constraint further increases the total repulsive interaction by about 10\%. The strength of the potential is of the order of $1\, k_BT$ at small separations $r$. The number of rejected conformations increases by over 50\% compared to only taking into account excluded volume interactions. 

The analysis of conformational properties of polymers brought close together reveals effects of the ring structure  both on size and shape. Ring polymers in proximity become swollen (Fig.~\ref{fig:dim:rgyr}) and strongly elongated (Fig.~\ref{fig:shape:ratios}) compared to the situation of isolated or far apart ones. While both effects are found for linear polymers, too, the effects are more pronounced for their ring counterparts. Additional effects from the non-catenation constraint concerning size and shape are visible, but small, indicating that the changes are mainly induced by the presence of additional material due to the rings being more compact than linear polymers. 

Remarkable effects of the ring structure are found concerning the alignment of ring polymers in proximity. There is a strong tendency to align perpendicular for short center-of-mass separations, an effect which is much more pronounced than for linear chains. At intermediate separations ($r\sim1-2 R_{g}$) a slightly parallel alignment is encountered. 

The transition from a linear to a ring polymer thus induces strong changes in the conformational and structural properties of a system. In fact, ring polymers adopt a much more ordered and regular state, showing less intermingling due to the increased repulsive interactions. The mutual alignment of ring polymers becomes much more different from a random state than found for linear polymers. 

Our findings suggest, that the ring topology of certain biopolymers like DNA and proteins, displays a benefit compared to the linear organization form. The formation of chromatin loops in higher eukaryotes therefore might play a dominant role concerning overall nuclear organizational principles. We have shown that loop structures lead to a strong effective repulsion and for a first time quantitatively analyzed the strength of the resulting interactions. Surely, the formation of multiple loops in the system of chromatin which has been proposed in several models~\cite{Grosberg1993, Bohn2007a, Mateos2009, Lieberman-Aiden2009} induces even stronger effects and therefore might be responsible for maintaining the segregated state of chromosomes found in several experiments~\cite{Cremer2001}. Indeed, studies of looping polymers have revealed an effect on the abundance of inter-chromosomal contacts~\cite{Cook2009, Bohn2009b}. Whereas it has already been proposed that the ring structure prevents chromosomes from entangling~\cite{Vettorel2009, Rosa2008}, such a topology also induces a much more ordered and aligned state than linear polymers. It remains to be shown, to which extend similar results are found for chromatin models with loops~\cite{Grosberg1993, Mateos2009}. We expect that both the  local behaviour (e.g. inside of transcription factories or heterochromatic regions) and the global behavior (segregation of chromosomes) might well be dominated by entropic and topological forces of chromatin loops. Contrary to a melt of ring polymers, which adopts a compact state, the dynamic formation of loops~\cite{Bohn2009b} might provide a versatile mechanism for the cell to switch between and maintain different local levels of compaction by facilitating or repressing loop formation.

\end{document}